\documentclass[aps,twocolumn,showpacs,superscriptaddress,groupedaddress, nofootinbib]{revtex4-1}

\usepackage{hyperref}
\usepackage{multirow}
\usepackage{mathrsfs,amsmath} 
\usepackage{verbatim}
\usepackage{amsmath}
\usepackage{amsfonts}
\usepackage{amssymb}
\usepackage{amsthm}
\usepackage{bm}
\usepackage{xparse}
\usepackage{xcolor}
\usepackage{textcase}
\usepackage{url}
\usepackage{lipsum}
\usepackage{appendix}
\usepackage{dsfont}
\usepackage{epstopdf}
\usepackage{footnote}
\usepackage{tgbonum}
\usepackage{blindtext}
\usepackage{enumitem}
\usepackage{mathrsfs,amsmath} 
\usepackage{float}
\usepackage{subfloat}
\usepackage{wrapfig}
\usepackage{pslatex}
\usepackage{amsmath}
\usepackage{amsfonts}
\usepackage{setspace}
\usepackage{epstopdf}
\usepackage{wrapfig}
\usepackage{csquotes}
\usepackage{hyperref}
\usepackage{graphicx}
\usepackage{amssymb}
\usepackage{multirow}
\usepackage{xcolor}
\usepackage{framed}
\usepackage{mathtools}
\usepackage{tcolorbox}

\definecolor{shadecolor}{rgb}{0.8,0.9,1}
\newcommand{\ket}[1]{| {#1} \rangle} 
\DeclareDocumentCommand{\Tr}{m m O{\big}}{{\rm Tr}_{\:\!{#1}}#3({#2}#3)}

\begin{document}
\title{Many Worlds are irrelevant for the problem of the arrow of time}
\author{Veronika Baumann} \author{Flavio Del Santo}
\affiliation{
Institute for Quantum Optics and Quantum Information (IQOQI-Vienna), A-1090 Vienna, Austria; and
Faculty of Physics, University of Vienna, A-1090 Vienna, Austria}
\altaffiliation[]{The authors contributed equally to this work.}
\date{\today}

\begin{abstract}
In a recent paper, S. Gao has claimed that, under the assumption that the initial state of the universe is a pure quantum state, only the many worlds interpretation can account for the observed arrow of time. We show that his argument is untenable and that if endorsed it potentially leads to undermine the search for a scientific explanation of certain phenomena.
\end{abstract}

\maketitle

\section{The past hypothesis}
\label{Introduction}


It is generally known that, despite the fact that both in classical (also relativistic) and quantum mechanics the equations of motion are invariant under time-reversal, we perceive a clear asymmetry between past and future and time seems to point toward a direction, the so-called \emph{arrow of time} \cite{eddington2019nature}. This is customarily explained in statistical terms by the fact that, although the motion of the single particles remain fully reversible, an increase in the macroscopic quantity entropy has a higher chance to occur (second law of thermodynamics).\footnote{Note that alternative explanation in terms of modifications or reinterpretations of classical physics \cite{del2019physics, e23101326} and of quantum physics (such as objective collapse theories \cite{PhysRevD.34.470}) can also explain time asymmetry in terms of fundamental irreversibly of physical process at the microscopic level.} More precisely, the second law asserts that  if at time $t_1$ the Boltzmann entropy $S(t_1)$ of a system is low, then at some later time $t_2 >t_1$, the probability that the entropy $S_2=S(t_2)>S(t_1)=S_1$ is high (close to 1), \emph{provided} that initially the entropy was even lower. It should be noted that this has the form of a conditional probabilistic law, namely it reads 
\begin{equation}
P\left(S_2>S_1|S_0\ll S_{max}\right)\approx 1, 
\label{2ndLaw}
\end{equation}
where 
$S_0$ is the entropy of the initial state of the system (at time $t_0$) and $S_{max}$ is the maximum entropy corresponding to the system reaching the thermal equilibrium. Hence --as noted by J. Earman, among others-- assuming a mere statistical validity of the second law of thermodynamics does not fully explain the problem of the arrow of time away, for ``the grounding of the relevant temporal asymmetries in entropic behavior observed today is to be found in the fact (posit?) that the early universe was in a low entropy state (the `Past Hypothesis').'' \cite{earman2006past}. (See also \cite{albert2001time} for a thorough discussion). The `Past hypothesis', however, seems problematic, since such an initial low-entropy state of the universe is very unlikely \cite{albert2001time,price1996time,wallace2011logic},
\begin{equation}
P\left(S_0\ll S_{max}\right)\approx 0, 
\label{P_count}
\end{equation}
and, hence, incompatible with typicality arguments concerning the beginning of our universe.
When applied to \emph{our}  universe the observed arrow of time means it must have been in a low entropy state at the beginning:
\begin{equation}
\forall t_n > t_m : S_n >S_m \implies S_0 \ll S_{max}.
\end{equation}
This is the same as to say that the probability that our universe started in an initial state with low entropy is very high:
\begin{equation}
P\left(S_0\ll S_{max} | S_n > S_m\text{,\ \ } \forall t_n > t_m  \right)\approx 1. 
\label{P_time}
\end{equation}
While this might create a sense of unease, it can for example be argued via some form of ``anthropic principle'': possible initial configurations of our universe that had not a low enough entropy (i.e. distant enough from the equilibrium), would have not allowed the formation of complex thermodynamic systems (like living beings) that we observe today.

Note that everything stated so far is known, has nothing to do with quantum theory and can be addressed without resorting to the latter. For instance, a simple classical thermodynamic explanation of the past hypothesis states that, under the assumption of ergodicity, there is a time at which the microscopic configurations will actuate any macro-state (including one with arbitrarily low entropy). Poincaré's recurrence theorem then ensures that, given a sufficiently long time, the system will be found in a arbitrarily small neighborhood of said state, leading to a recursive arrow of time (see \cite{earman2006past} for a detailed discussion).

\section{Comment on ``Time's arrow points to many worlds''}
\label{Reply}

In a recent preprint \cite{gao2021time}, however, S. Gao has claimed that  the ``the past hypothesis is not necessary, and the initial state of the universe may be a general superposition of both low-entropy states and high-entropy states. In this case, the many-worlds interpretation of quantum mechanics (MWI) can still account for the thermodynamic arrow of time, although the single-world quantum theories cannot.'' \cite{gao2021time}. While the first part --that the initial state of the universe is in a quantum superposition-- is potentially arguable, the latter is clearly false, as we will argue below.

Gao's initial assumption seems to be that since (i) quantum physics is our best theory of mechanics and (ii) there are no limits of validity to the domain of application of quantum theory (in terms of scales of mass, energies, number of particles, etc.), the initial state of the universe should be a quantum state. He considers a general pure state at the initial time $t_0$ to be in a quantum superposition of all possible macroscopic states with a well-defined entropy, i.e.,
\begin{equation}
\ket {\psi(t_0)} = \sum_i  a_i \ket{\psi^L}_i + \sum_j  b_j \ket{\psi^H}_j, 
\label{state}
\end{equation}
where $\ket{\psi^L}_i$ and $\ket{\psi^H}_j$ are, respectively, low entropy macro-states (that would lead to the observed arrow of time) and high entropy macro-states (that would not lead to the observed arrow of time), while $a_i$ and $b_j$ are the associated coefficient that fulfill the normalization condition. A typicality argument implies that $\sum_i |a_i|^2 \ll \sum_j |b_j|^2$, since low entropy states are much less likely among the possible macroscopic configurations.

Gao then proceeds to claim that single-world interpretations of quantum theory --either with unitary evolution (like Bohmian mechanics) or objective collapse theories-- require that our universe was initially in a high entropy state and continues to be so. This means that they cannot account for our observation of the arrow of time. Furthermore, adopting a many-worlds interpretation of quantum theory supposedly would imply that our universe (thought as one branch of the multiverse given by equation \eqref{state}) may have started in a low entropy initial state and, hence, we observe an arrow of time. In the author's words, ``the key is to notice that single-world quantum theories predict that our universe is typical, and it evolves from a high-amplitude decoherent branch of the initial universal wave function which has a large squared amplitude. While MWI predicts that our universe may be atypical, and it may evolve from a low-amplitude decoherent branch of the initial
universal wave function which has a very small squared amplitude.'' \cite{gao2021time}.

The claim about single-world interpretations is based on the initial assumption that $\sum_i |a_i|^2 \ll \sum_j |b_j|^2$ (typicality) which can be viewed as the quantum analog of equation \eqref{P_count}. Note that the same is, of course, true in a MWI where the number of universes starting in a high entropy macro-state is much higher than that of those starting in a low entropy macro-state.

In his article, Gao further states that the probability of our universe starting out in a low entropy state in a MWI ``may be equal to one''. This is not justified further in the main text but was later clarified to refer to the conditional probability in equation \eqref{P_time}, justified  as to be ``similar to the reasoning of the quantum suicide thought experiment'' \cite{Gao_email}. The latter, to us, appears to be essentially an argument based on (a quantum version of) the anthropic principle. Since this probability assignment is \emph{conditioned on the fact that we do observe an arrow of time}, the same assignment can be made within a single-world interpretation using arguments similar to the classical ones stated in section \ref{Introduction}.

The seeming advantage of a many-worlds interpretation is based on allowing to condition the probability of the initial state of our universe on the fact that we do observe an arrow of time, while disallowing to do so for single-world interpretations, in practice unfairly comparing the two different probabilities given by equations \eqref{P_count} and \eqref{P_time}, respectively. If one accepts the proposed initial state of the universe as given by equation \eqref{state}, \emph{any} interpretation of quantum theory would state that our observed universe has an overwhelmingly low probability amplitude. This probability is to be understood as a measure of likelihood  for our universe to have started in a particular state among all potentially conceivable initial configurations. However, given that we do observe an arrow of time, our universe (either to be understood as the only existing universe, or as a branch of the multiverse) must have started in a low entropy initial state.

In conclusion, while Gao manages to rephrase the problem of the past hypothesis for a MWI, this, contrary to his claim, does not constitute evidence that a many worlds theory is empirically favored by our experiencing an arrow of time. Said claim is not only wrong but we regard it as inherently problematic. The author of \cite{gao2021time}, in fact, seems to draw more general conclusions from it, when suggesting that ``besides the thermodynamic arrow of time, the matter-antimatter asymmetry may be another example'' of ``evidence'' for many worlds. However, by the same line of argument \emph{every} atypicality at the cosmological scale (i.e. where certain parameters or boundary conditions, independently of how unlikely they may be, should be fine-tuned to fit with observations) would just ``strongly support MWI and disfavor the single-world quantum theories'' \cite{gao2021time} rather than requiring a scientific explanation. This appears to be detrimental to scientific inquiry.\\ \\
\textit{Acknowledgments} -- F.D.S. acknowledges support from European Innovation Council and SMEs Executive Agency, TEQ 766900 / Project OEUP0259. V.B. acknowledges funding from FQXi.

\bibliographystyle{unsrt}
\bibliography{Refs}

\end{document}